\documentclass[12pt]{article}
\usepackage{amsmath,amssymb}
\begin{document}

\begin{titlepage}
\begin{center}

{\Large \bf Einstein Equations from Riemann-only Gravitational Actions}

\bigskip

\bigskip

{\bf  Durmu{\c s} Demir$^{a}$, Oktay Do{\~g}ang{\"u}n$^b$, Tongu{\c c} Rador$^{c}$, Selin Soysal$^{d}$}\\
\smallskip

{ \small \it  
$^a$ Department of Physics,
{\.I}zmir Institute of Technology, IZTECH, TR35430, {\.I}zmir, Turkey\\
$^{b}$ Department of Physical Sciences, University of Naples \& INFN, Naples, Italy\\
$^{c}$ Department of Physics, Bo{\u g}azi{\c c}i University,
Bebek TR34342, {\.I}stanbul, Turkey\\
$^{d}$ Department of Mathematics,
{\.I}zmir Institute of Technology, IZTECH, TR35430, {\.I}zmir, Turkey}

\bigskip

\vspace*{.5cm}

{\bf Abstract}\\
\end{center}
\noindent
Here show that, pure affine actions based solely on the Riemann curvature
tensor lead to Einstein field equations for gravitation. The matter and
radiation involved are general enough to impose no restrictions on 
material dynamics or vacuum structure. This dynamical equivalence to 
General Relativity can be realized via also diffeomorphism breaking.
\bigskip

\bigskip

\end{titlepage}

Einstein field equations of gravitation locally relate Ricci curvature tensor to energy-momentum tensor -- a symmetric
tensor field $T_{\alpha\beta}$ comprising energy density, momentum density, momentum flux, pressure and shear stress of matter, radiation and vacuum. 
The Ricci tensor, however, gives only limited information about the spacetime curvature because the grand curvature of spacetime is encoded in 
the Riemann tensor not in its trace -- the Ricci tensor. The exception occurs in $D=3$ dimensions in which Einstein field equations completely 
determine the spacetime curvature simply because the Ricci and Riemann tensors comprise the same degrees of freedom. For $D>4$, however, 
this is not the case since Riemann tensor comprises much more curvature components than Ricci tensor. In General Relativity (GR), the sole geometrodynamical 
variable is the metric tensor $g_{\alpha\beta}$, which naturally  defines a connection
\begin{eqnarray}
\label{LC}
\gamma^{\mu}_{\alpha\beta} = \frac{1}{2} g^{\mu\nu} \left(\partial_{\alpha} g_{\beta\nu} + \partial_{\beta} g_{\nu\alpha} - \partial_{\nu} g_{\alpha\beta}\right)
\end{eqnarray}
through which all curvature tensors are expressed in terms of the metric tensor. Therefore, the Einstein field equations determine 
only the metric tensor with which curvature tensors are computed {\it a posteriori}. This is best exemplified by the Schwarzschild solution \cite{schwarzschild}
where the Ricci tensor vanishes identically yet it yields a metric tensor for which the Riemann tensor does not vanish.

In the present work, our goal is to reveal dynamics of the grand curvature of spacetime in connection with Einstein field 
equations for gravitation. In particular,  we will explore the geometrodyncmical processes which induce the metric 
tensor, necessitate the Levi-Civita connection, and bring in the energy-momentum tensor. For this purpose we shall 
construct invariant actions involving only the grand curvature of spacetime encoded in Riemann tensor, and analyze 
the dynamical equations they yield in physically relevant cases. It will be seen that Einstein field equations
dynamically emerge as the consistent equations. 

It is easy to see that the said goal cannot be achieved in a spacetime endowed with metric. The reason is that the relation
\begin{eqnarray}
\label{decompose}
R_{\mu \alpha \nu \beta}\left(\gamma\right) &=& W_{\mu \alpha \nu \beta}\left(\gamma\right) + \frac{1}{2} \left( g_{\mu\nu} R_{\beta\alpha}\left(\gamma\right) 
- g_{\mu\beta} R_{\nu \alpha}\left(\gamma\right) \right) + \frac{1}{2} \left( g_{\alpha\beta} R_{\nu\mu}\left(\gamma\right) - g_{\alpha\nu} R_{\beta\mu}\left(\gamma\right)\right)\nonumber\\ &-& 
\frac{1}{6} R\left(g,\gamma\right) \left( g_{\mu\nu} g_{\beta\alpha} - g_{\mu\beta} g_{\nu\alpha}\right)
\end{eqnarray}
decomposes Riemann tensor in terms of the Weyl tensor $W_{\mu \alpha \nu \beta}\left(\gamma\right)$, the Ricci tensor $R_{\mu\nu}\left(\gamma\right) = 
g^{\alpha\beta} R_{\mu \alpha \nu \beta}\left(\gamma\right)$, and the Ricci scalar $R\left(g,\gamma\right) = g^{\mu\nu} R_{\mu\nu}\left(\gamma\right)$ so that  
the sought dynamical equation for the Riemann tensor is determined by Einstein field equations up to a traceless piece contained in the Weyl tensor. 
Therefore, for structuring an independent dynamical equation involving Riemann tensor the spacetime must be endowed with no metrical structure to start with yet metric 
and dynamical quantities like energy-momentum tensor can arise {\it a posteriori} dynamically (This affine spacetime has been introduced
by Eddington \cite{eddington}, and further explored by Einstein \cite{einstein} and Schroedinger \cite{schroedinger}. See the reveiw volumes \cite{review}.). 
In view of this, we consider a spacetime manifold where notions of length and angle are absent to start with. The manifold is 
then characterized by a single geometrodynamical variable -- the affine connection $\Gamma$. It is an acceleration field which governs geodesics 
of test masses, and completely determines the grand curvature of spacetime encoded in Riemann tensor
\begin{eqnarray}
\label{riemann}
{\mathfrak{R}}^{\mu}_{\alpha\nu\beta}\left(\Gamma\right) = \partial_{\nu} \Gamma^{\mu}_{\beta\alpha} - \partial_{\beta} \Gamma^{\mu}_{\nu\alpha}
+\Gamma^{\mu}_{\nu\lambda} \Gamma^{\lambda}_{\beta\alpha} - \Gamma^{\mu}_{\beta\lambda} \Gamma^{\lambda}_{\nu\alpha}
\end{eqnarray}
as well as the partial curvature of spacetime inscribed in Ricci tensor
\begin{eqnarray}
\label{ricci}
{\mathcal{R}}_{\alpha\beta}\left(\Gamma\right) = \partial_{\mu} \Gamma^{\mu}_{\beta\alpha} - \partial_{\beta} \Gamma^{\mu}_{\mu\alpha}
+\Gamma^{\mu}_{\mu\lambda} \Gamma^{\lambda}_{\beta\alpha} - \Gamma^{\mu}_{\beta\lambda} \Gamma^{\lambda}_{\mu\alpha}\,.
\end{eqnarray}
Here we specialized to the symmetric connection $\Gamma^{\mu}_{\alpha\beta} = \Gamma^{\mu}_{\beta\alpha}$ for which the spacetime is
not twirled, only curved. Having no metric tensor, the curvature tensors (\ref{riemann}) and (\ref{ricci})  cannot be contracted 
any further to obtain invariants (such as the Ricci scalar in GR). The absence of invariants, however, does not prohibit construction of invariant actions because
differential volume element 
\begin{eqnarray}
d^4 x = \epsilon_{\alpha_0 \alpha_1 \alpha_2 \alpha_3 } dx^{\alpha_0} \wedge dx^{\alpha_1} \wedge dx^{\alpha_2} \wedge dx^{\alpha_3}
\end{eqnarray}
is a scalar density not a true scalar since  it transforms as $d^4 x = {\texttt{Det}}\left[\frac{\partial x}{\partial x^{\prime}}\right] d^{4} x^{\prime}$ 
under a coordinate transfomation $x^{\mu} \rightarrow x^{\prime\, \mu}$. Therefore, all one needs is a scalar density ${\mathcal{L}}$
such that the product $d^{4}x\ {\mathcal{L}}$ is invariant. The requisite ${\mathcal{L}}$, which must transform inversely to $d^{4} x$,
must obviously be related to determinants of tensor fields. Our goal is to form a dynamical equation that directly involves the grand
curvature of spacetime. Therefore, the scalar density ${\mathcal{L}}$ must involve solely the Riemann tensor since it is only
this curvature tensor that sensitively probes all components of the grand curvature of spacetime. Consequently, ${\mathcal{L}}$ can be expanded as 
\begin{eqnarray}
\label{density}
{\mathcal{L}} = \sum_i a_i \left( {\mbox{Det}}\left[\left({\mathfrak{R}}\odot{\mathfrak{R}}\right)_i\right]\right)^{1/4}
+ \sum_i b_i \left( {\mbox{Det}}\left[\left({\mathfrak{R}}\odot{\mathfrak{R}}\odot{\mathfrak{R}}\right)_i\right]\right)^{1/6} + \cdots
\end{eqnarray}
each term of which involves $i=1,2,\dots$ distinct contractions of Riemann tensors. For instance, $\left({\mathfrak{R}}\odot{\mathfrak{R}}\right)_i$ 
is a rank-4 tensor field with possible forms
\begin{eqnarray}
\label{forms}
\left[\left({\mathfrak{R}}\odot{\mathfrak{R}}\right)_{i}\right]_{\mu\alpha\nu\beta} \in \left\{
{\mathcal{R}}_{\alpha\beta} {\mathcal{R}}_{\mu\nu},  {\mathfrak{R}}^{\rho}_{\alpha\kappa\beta}{\mathfrak{R}}^{\kappa}_{\mu\rho\nu}, 
{\mathfrak{R}}^{\rho}_{\kappa\alpha\beta}{\mathfrak{R}}^{\kappa}_{\mu\rho\nu},
{\mathfrak{R}}^{\rho}_{\kappa\alpha\beta}{\mathfrak{R}}^{\kappa}_{\rho\mu\nu}\right\}
\end{eqnarray}
and determinant
\begin{eqnarray}
{\mbox{Det}}\left[\left({\mathfrak{R}}\odot{\mathfrak{R}}\right)_i\right] &=& \frac{1}{\left(4 !\right)^2}\epsilon^{\mu_0 \mu_1 \mu_2 \mu_3} \epsilon^{\alpha_0
\alpha_1 \alpha_2 \alpha_3} \epsilon^{\nu_0 \nu_1 \nu_2 \nu_3} \epsilon^{\beta_0 \beta_1 \beta_2 \beta_3}
\left[\left({\mathfrak{R}}\odot{\mathfrak{R}}\right)_{i}\right]_{\mu_0\alpha_0\nu_0\beta_0}\nonumber\\ &\times&
\left[\left({\mathfrak{R}}\odot{\mathfrak{R}}\right)_{i}\right]_{\mu_1\alpha_1\nu_1 \beta_1} 
\left[\left({\mathfrak{R}}\odot{\mathfrak{R}}\right)_{i}\right]_{\mu_2\alpha_2\nu_2\beta_2}
\left[\left({\mathfrak{R}}\odot{\mathfrak{R}}\right)_{i}\right]_{\mu_3\alpha_3\nu_3\beta_3}
\end{eqnarray}
which is a generalization of the usual determinant to rank-4 tensors \cite{deter}. The determinants of rank-6 and
higher tensors in (\ref{density}) involve contractions with 6 and higher $\epsilon^{\mu_0 \mu_1 \mu_2 \mu_3}$ symbols.

The scalar density in (\ref{density}) induces a Riemann-only invariant action
\begin{eqnarray}
\label{action}
I\left[\Gamma\right] = \int d^{4}x {\mathcal{L}}
\end{eqnarray}
which behaves stationary against the variation in curvature
\begin{eqnarray}
\label{variation}
\delta {\mathfrak{R}}^{\mu}_{\alpha\nu\beta} = \nabla_{\nu}\Big( \delta\Gamma^{\mu}_{\beta\alpha}\Big) -
\nabla_{\beta}\Big(\delta\Gamma^{\mu}_{\nu\alpha}\Big)
\end{eqnarray}
if physically correct curvature configurations are attained. This very stationarity property
\begin{eqnarray}
\delta I\left[\Gamma\right] = 0
\end{eqnarray}
locally translates into the equations of motion
\begin{eqnarray}
\label{eom-riemann}
\nabla_{\nu} \left(\frac{\delta {\mathcal{L}}}{\delta {\mathfrak{R}}^{\mu}_{\alpha\nu\beta}}  \right) = 0 
\end{eqnarray}
which possess a nontrivial solution
\begin{eqnarray}
\label{soln-riemann}
\frac{\delta {\mathcal{L}}}{\delta {\mathfrak{R}}^{\mu}_{\alpha\nu\beta}} = {\mathfrak{T}}^{\beta\nu\alpha}_{\quad \mu}
\end{eqnarray}
provided that ${\mathfrak{T}}^{\beta\nu\alpha}_{\quad \mu}$ is a divergence-free tensor density which sources the grand curvature of spacetime. The problem is to determine this
non-geometrical source term. If the equation of motion (\ref{eom-riemann}) were a vanishing gradient then its integral ${\mathfrak{T}}^{\beta\nu\alpha}_{\quad \mu}$ would involve 
only the metric tensor $g_{\alpha\beta}$. The reason is that equation of motion forces $\Gamma^{\mu}_{\alpha\beta}$ to  be compatible with 
${\mathfrak{T}}^{\beta\nu\alpha}_{\quad \mu}$, which can always be decomposed in terms of a metric field $g_{\alpha\beta}$. Nevertheless, the equation of motion
(\ref{eom-riemann}) is a vanishing divergence, not a vanishing gradient, and its integral  ${\mathfrak{T}}^{\beta\nu\alpha}_{\quad \mu}$ necessarily involves novel
degrees of freedom beyond the metric tensor. Clearly, ${\mathfrak{T}}^{\beta\nu\alpha}_{\quad \mu}$  must be made up of at least two independent tensor fields such that while one of them provides 
the determinant factor needed for density property, the other takes care of the tensorial structure. More explicitly, in view of the structure of ${\mathcal{L}}$, 
the tensor density ${\mathfrak{T}}^{\beta\nu\alpha}_{\quad \mu}$  must be of the form
\begin{eqnarray}
\label{soln-dav}
{\mathfrak{T}}^{\beta\nu\alpha}_{\quad \mu}\sim \left({\texttt{Det}}\left[{\mathcal{M}}\right]\right)^{1/4}\, {\mathcal{N}}^{\beta\nu\alpha}_{\quad \mu}
\end{eqnarray}
where ${\mathcal{M}}$ is a rank-four tensor, and there are at least two independent structures contained in ${\mathcal{M}}$ and ${\mathcal{N}}$. Physically, it is
advantageous to express ${\mathcal{M}}$ wholly in terms of the metric tensor. To determine its structure, it suffices to go to the vanishing gradient limit in which
case ${\mathcal{M}}$ and $\mathcal{N}$ become identical, and symmetries of the left-hand side of (\ref{eom-riemann}) lead us uniquely to 
\begin{eqnarray}
\label{M-tensor}
{\mathcal{M}}_{\mu\alpha\nu\beta} = g_{\mu\nu} g_{\beta\alpha} - g_{\mu\beta} g_{\nu\alpha}
\end{eqnarray} 
as an appropriate choice (up to a multiplicative constant that can be absorbed in ${\mathcal{N}}^{\beta\nu\alpha}_{\quad \mu}$). 
A straightforward calculation gives ${\texttt{Det}}\left[{\mathcal{M}}\right] = 4 g^2$, where we defined 
$g = {\texttt{Det}}\left[g_{\alpha\beta}\right]$.
 
Having ${\mathcal{M}}$ fixed as in (\ref{M-tensor}),  it is necessary to exercise an independent structure for ${\mathcal{N}}^{\beta\nu\alpha}_{\quad \mu}$. In general, 
structure of ${\mathcal{N}}^{\beta\nu\alpha}_{\quad \mu}$ is governed by the symmetries of (\ref{eom-riemann}), and it uniquely leads to the tensor density
\begin{eqnarray}
\label{soln2-riemann}
{\mathfrak{T}}^{\beta\nu\alpha}_{\quad \mu} = \sqrt{2 g} \Big\{ {\mathcal{T}}^{\beta\nu\alpha}_{\quad \mu}
+ c_1 \big( {\mathcal{T}}^{\beta\alpha} \delta^{\nu}_{\mu} - \delta^{\beta}_{\mu} {\mathcal{T}}^{\nu\alpha}\big)
+ c_2 \big( g^{\beta\alpha} {\mathcal{T}}^{\nu}_{\mu} - {\mathcal{T}}^{\beta}_{\mu} g^{\nu\alpha}\big)
+ c_3 {\mathcal{T}} \big( g^{\beta\alpha} \delta^{\nu}_{\mu} - \delta^{\beta}_{\mu} g^{\nu\alpha}\big)  \Big\}
\end{eqnarray}
where $c_i$ are constants, and ${\mathcal{T}}^{\beta\nu\alpha}_{\quad \mu}$ is meant to be not entirely expressible in terms  of the metric tensor. In this expansion,
${\mathcal{T}}^{\beta\alpha} = {\mathcal{T}}^{\beta\nu\alpha}_{\quad \nu}$, ${\mathcal{T}}^{\nu}_{\mu} = g_{\beta\alpha} {\mathcal{T}}^{\beta\nu\alpha}_{\quad \mu}$
and ${\mathcal{T}}=g_{\beta\alpha} {\mathcal{T}}^{\beta\alpha}$ are the lower-rank structures deriving from ${\mathcal{T}}^{\beta\nu\alpha}_{\quad \mu}$. The
main constraint is that (\ref{soln2-riemann}) must have vanishing divergence, and this is impossible to satisfy
unless divergence of ${\mathcal{T}}^{\beta\nu\alpha}_{\quad \mu}$ is related linearly to gradients of its lower-rank contractions.
More explicitly, ${\mathcal{T}}^{\beta\nu\alpha}_{\quad \mu}$ must enjoy a relation of the form
\begin{eqnarray}
\label{diverg0}
\nabla_{\nu} {\mathcal{T}}^{\beta\nu\alpha}_{\quad \mu} =  c_4 \nabla_{\mu} {\mathcal{T}}^{\beta\alpha} + c_5 \nabla^{\alpha} {\mathcal{T}}^{\beta}_{\mu} + 
c_6 \nabla^{\beta} {\mathcal{T}}^{\alpha}_{\mu}
\end{eqnarray}
whose contraction 
\begin{eqnarray}
\label{diverg}
\nabla_{\nu} {\mathcal{T}}^{\nu}_{\mu} = \frac{c_4}{1-c_5-c_6} \nabla_{\mu} {\mathcal{T}}
\end{eqnarray}
provides an important link between the divergence and gradient of ${\mathcal{T}}$ tensors of different ranks. With these two 
crucial properties of ${\mathcal{T}}^{\beta\nu\alpha}_{\quad \mu}$, the divergence of ${\mathfrak{T}}^{\beta\nu\alpha}_{\quad \mu}$ vanishes identically
if the expansion coefficients in (\ref{soln2-riemann}) take the specific values
\begin{eqnarray}
\label{coeffs}
c_2 = -c_4 = c_5 = c_1,\, c_3 = \frac{c_1^2}{1-c_1},\, c_6 = 0.
\end{eqnarray}

Having explicitly constructed the divergence-free density ${\mathcal{T}}^{\beta\nu\alpha}_{\quad \mu}$, one notices that the 
relation (\ref{diverg}) has far-reaching consequences because it guarantees the existence of a divergence-free tensor field ${\mathbb{T}}_{\alpha\beta}$ ({\it i. e.} 
$\nabla_{\alpha} {\mathbb{T}}^{\alpha\beta} = 0$) such that
\begin{eqnarray}
{\mathcal{T}}_{\alpha\beta} = {\mathbb{T}}_{\alpha\beta} - \frac{c_1}{3 c_1 + 1} {\mathbb{T}} g_{\alpha\beta}
\end{eqnarray}
where uses have been made of the solution (\ref{coeffs}) and of the definition ${\mathbb{T}} = g^{\mu\nu} {\mathbb{T}}_{\mu\nu}$.

The solution for ${\mathcal{T}}^{\beta\nu\alpha}_{\quad \mu}$ in (\ref{soln2-riemann}) immediately fixes the affine connection $\Gamma^{\mu}_{\alpha\beta}$ to 
be the Levi-Civita connection $\gamma^{\mu}_{\alpha\beta}$ defined in (\ref{LC}). This arises automatically by the solution (\ref{M-tensor}) 
for ${\mathcal{M}}_{\mu\alpha\nu\beta}$. Everywhere in the following thus $\Gamma^{\mu}_{\alpha\beta}$ will be replaced by $\gamma^{\mu}_{\alpha\beta}$.

For the values of $c_i$ found in (\ref{coeffs}), the tensor density ${\mathfrak{T}}^{\beta\nu\alpha}_{\quad \mu}$ becomes 
\begin{eqnarray}
\label{soln2p-riemann}
{\mathfrak{T}}^{\beta\nu\alpha}_{\quad \mu} &=& \sqrt{2 g} \Big\{ {\mathcal{T}}^{\beta\nu\alpha}_{\quad \mu}\nonumber\\
&+& c_1 \big( {\mathcal{T}}^{\beta\alpha} \delta^{\nu}_{\mu} - \delta^{\beta}_{\mu} {\mathcal{T}}^{\nu\alpha}\big)
+ c_1 \big( g^{\beta\alpha} {\mathcal{T}}^{\nu}_{\mu} - {\mathcal{T}}^{\beta}_{\mu} g^{\nu\alpha}\big)
+ \frac{c_1^2}{1-c_1} {\mathcal{T}} \big( g^{\beta\alpha} \delta^{\nu}_{\mu} - \delta^{\beta}_{\mu} g^{\nu\alpha}\big) \Big\}
\end{eqnarray}
whose second line obtains a similar form as the Ricci-dependent parts of the decomposition (\ref{decompose}) 
(this similarity becomes identity for $c_1= -{1}/{2}$). This implies that ${\mathcal{T}}_{\alpha\beta}$, a non-geometrical dynamical 
tensor field, might be linearly related to the Ricci curvature tensor. Also implied 
is that the left-hand side of (\ref{soln-riemann}) must be proportional to the Riemann tensor. Having dynamically reached a relation 
mimicking the decomposition of Riemann tensor in (\ref{decompose}) is an important step in making sense of the 
Riemann-only actions like (\ref{action}). These said relations must not be taken as implying a replacement
rule for Ricci tensor and others. The reason is that neither (\ref{diverg0}) nor (\ref{diverg}) is identical to 
corresponding Bianchi identities satisfied by Riemann and Ricci tensors. This property ensures that ${\mathcal{T}}^{\beta\nu\alpha}_{\quad \mu}$ tensor is 
not a geometrical field enjoying the Jacobi identity obeyed by the connection $\Gamma$. In other words, as has been 
emphasized from the beginning, ${\mathcal{T}}^{\beta\nu\alpha}_{\quad \mu}$ is indeed a non-geometrical tensor 
field sourcing the spacetime curvature. In fact, in view of Einstein field equations in GR, the said 
correspondance between Ricci tensor and ${\mathcal{T}}_{\alpha\beta}$ implies that the latter can be 
related to the energy-momentum tensor of matter, radiation and vacuum. As noted above, the second
line of (\ref{soln2p-riemann}) becomes identical to the Ricci-part of (\ref{decompose}) for $c_1=-1/2$.
This means that $c_1=-1/2$ sets the geometrical regime in which the derivative identities (\ref{diverg0})
and (\ref{diverg}) satisfied by ${\mathcal{T}}^{\beta\nu\alpha}_{\quad \mu}$ reduce to Bianchi 
identities. However, a closer look at (\ref{soln2-riemann}) reveals that the trace ${\mathcal{T}}^{\beta\mu\alpha}_{\quad \mu}$
identically vanishes for $c_1=-1/2$  because it is proportional to $(1+ 2 c_1)$. This observation guarantees thus
that there is no geometrical regime for the tensor density ${\mathcal{T}}^{\beta\nu\alpha}_{\quad \mu}$, which serves 
as a dynamical source field for grand curvature of spacetime.

Having determined the source term ${\mathcal{T}}^{\beta\nu\alpha}_{\quad \mu}$, it remains to specify the gravitational action itself.
It is with determination of ${\mathcal{L}}$ that one can obtain the dynamical equations for gravitation by using the solution (\ref{soln-riemann}).
As far as the equations of motion are concerned, it physically suffices to stop at the rank-4 level of the expansion (\ref{density}). 
Furthermore, one may, for simplicity, content with just one of the  terms in (\ref{forms}) such as
\begin{eqnarray}
\label{trial}
a_2 \left( {\mbox{Det}}\left[{\mathfrak{R}}^{\rho}_{\alpha\kappa\beta}{\mathfrak{R}}^{\kappa}_{\mu\rho\nu}
\right]\right)^{1/4}
\end{eqnarray} 
whose direct analysis shows that it indeed leads to an equation for Riemann tensor. However, because of the identities
(\ref{diverg0}) and (\ref{diverg}), the Riemann tensor turns out to possess vanishing divergence. This is an unacceptable
solution as it violates Bianchi identities. The lesson from this result is that one must pick up at least two 
distinct terms in the expansion (\ref{density}). For definiteness, adding also the first term in (\ref{forms}) to (\ref{trial}) 
one obtains a more general action density
\begin{eqnarray}
\label{lagran}
{\mathcal{L}} = a_1 \left( {\mbox{Det}}\left[ {\mathcal{R}}_{\alpha\beta} {\mathcal{R}}_{\mu\nu}\right]\right)^{1/4}
 + a_2 \left( {\mbox{Det}}\left[{\mathfrak{R}}^{\rho}_{\alpha\kappa\beta}{\mathfrak{R}}^{\kappa}_{\mu\rho\nu}
\right]\right)^{1/4}
\end{eqnarray}
which gives the equation of motion
\begin{eqnarray}
\label{soln2pp-riemann}
a_2 \left( {\mbox{Det}}\left[\left({\mathfrak{R}}\odot{\mathfrak{R}}\right)_2\right]\right)^{1/4} 
\left({\mathfrak{R}}^{-1}\right)^{\beta\nu\alpha}_{\quad \mu}
+ a_1 \left( {\mbox{Det}}\left[{\mathcal{R}}\right]\right)^{1/2} 
\left( \delta^{\nu}_{\mu} \left({\mathcal{R}}^{-1}\right)^{\beta\alpha}
-  \left({\mathcal{R}}^{-1}\right)^{\nu\alpha}\delta^{\beta}_{\mu}\right) = 2 {\mathfrak{T}}^{\beta\nu\alpha}_{\quad \mu}
\end{eqnarray}
In extracting curvature tensors from this equation, one notices that each one of the two terms at the left-hand side,
the inverse Riemann and Ricci pieces, can always be expanded as in (\ref{soln2-riemann}). The only difference is that vanishing
of divergence is no longer a requirement. As a matter of convenience, the Riemann part in (\ref{soln2pp-riemann}) can be expressed as
\begin{eqnarray}
\label{soln3-riemann}
a_2 \left( {\mbox{Det}}\left[\left({\mathfrak{R}}\odot{\mathfrak{R}}\right)_2\right]\right)^{1/4}
\left({\mathfrak{R}}^{-1}\right)^{\beta\nu\alpha}_{\quad \mu} &=& \sqrt{2 g}
\Big\{ {\mathcal{T}}^{\beta\nu\alpha}_{\quad \mu}
+ \tilde{c}_1 \big( {\mathcal{T}}^{\beta\alpha} \delta^{\nu}_{\mu} - \delta^{\beta}_{\mu} {\mathcal{T}}^{\nu\alpha}\big)\nonumber\\
&+& \tilde{c}_2 \big( g^{\beta\alpha} {\mathcal{T}}^{\nu}_{\mu} - {\mathcal{T}}^{\beta}_{\mu} g^{\nu\alpha}\big)
+   \tilde{c}_3 {\mathcal{T}} \big( g^{\beta\alpha} \delta^{\nu}_{\mu} - \delta^{\beta}_{\mu} g^{\nu\alpha}\big)  \Big\}
\end{eqnarray}
where we reiterate that the expansion coefficients $\tilde{c}_i$ here are meant not to take the specific values in (\ref{coeffs}).

Having equation of motion (\ref{soln-riemann}) solved by (\ref{soln2pp-riemann}) with the supplementary relation (\ref{soln3-riemann}), there
arise two distinct ways for extracting the Ricci tensor, which must satisfy Einstein field equations. In the first, one first solves for the Riemann tensor
\begin{eqnarray}
\label{soln4-riemann}
{\mathfrak{R}}^{\mu}_{\alpha\nu\beta}\left(\gamma\right) = \frac{\sqrt{2}}{a_2}\left[{\mathcal{T}}^{\mu}_{\alpha\nu\beta}
+ \tilde{c}_1 \big( {\mathcal{T}}^{\mu}_{\nu} g_{\alpha\beta} - \delta^{\mu}_{\beta} {\mathcal{T}}_{\alpha\nu}\big)
+ \tilde{c}_2 \big( \delta^{\mu}_{\nu} {\mathcal{T}}_{\alpha\beta} - {\mathcal{T}}^{\mu}_{\beta} g_{\alpha\nu}\big)
+ \tilde{c}_3  {\mathcal{T}} \big( \delta^{\mu}_{\nu} g_{\alpha\beta} - \delta^{\mu}_{\beta} g_{\alpha\nu}\big)\right]
\end{eqnarray}
from the defining relation (\ref{soln3-riemann}). In the second, one directly extracts the Ricci tensor
\begin{eqnarray}
\label{soln5-riemann}
{\mathcal{R}}_{\alpha\beta}\left(\gamma\right) = \frac{\sqrt{8}}{3 a_1}\left[\left(3 ({c}_1-\tilde{c}_1) - ({c}_2-\tilde{c}_2)\right) {\mathcal{T}}_{\alpha\beta}
+ \left( ({c}_2-\tilde{c}_2) + 3 ({c}_3-\tilde{c}_3)\right) {\mathcal{T}} g_{\alpha\beta}\right]
\end{eqnarray}
after replacing (\ref{soln3-riemann}) into (\ref{soln2pp-riemann}). The two results, (\ref{soln4-riemann}) and (\ref{soln5-riemann}), consistently
lead to the dynamical equation
\begin{eqnarray}
\label{solution-riemann-einstein}
{\mathcal{R}}_{\alpha\beta}\left(\gamma\right) = -\frac{\sqrt{8}(2 c_1+1)}{3 a_1 + 2 a_2}  \left( {\mathbb{T}}_{\alpha\beta} - \frac{1}{2} {\mathbb{T}} g_{\alpha\beta}\right)
\end{eqnarray}
if the couplings $\tilde{c}_i$ take the values
\begin{eqnarray}
\label{coeffs-riemann}
\tilde{c}_1 &=& \frac{1}{(12 a_1 + 8 a_2)} \left( 3 a_1 (1 + 6 c_1) - 2 a_2 (1 - 2 c_1)\right)\nonumber\\
\tilde{c}_2 &=& \frac{1}{(12 a_1 + 8 a_2)} \left( 3 a_1 (1 - 2 c_1) + 2 a_2 (3 + 2 c_1)\right)\nonumber\\
\tilde{c}_3 &=& \frac{1}{(12 a_1 + 8 a_2)(c_1-1)} \left( a_1 (1 + 5 c_1 - 6 c_1^2) - 2 a_2 (1  + c_2 + 2 c_1^2)\right)
\end{eqnarray}
as required by Bianchi identities and equivalence of the two Ricci tensors in (\ref{soln4-riemann}) and (\ref{soln5-riemann}).

Physically, ${\mathbb{T}}_{\alpha\beta}$ in (\ref{solution-riemann-einstein}) must be proportional to energy-momentum tensor 
$T_{\alpha\beta}$ since, in the Newtonian limit, gravitational potential $1+g_{00}$ is sourced by
${\mathbb{T}}_{00}$, which must give the local mass density. Therefore, on dimensional grounds, ${\mathbb{T}}_{\alpha\beta} \propto T_{\alpha\beta}/M^2$
where $M$ is a mass scale, and the gravitational field equations (\ref{solution-riemann-einstein}) can be recast as
\begin{eqnarray}
\label{eq-einstein}
{\mathcal{R}}_{\alpha\beta}\left(\gamma\right) =  8\pi G_N  \left( T_{\alpha\beta} - \frac{1}{2} T g_{\alpha\beta}\right)
\end{eqnarray}
where the coefficient $\displaystyle{-{\sqrt{8}(2 c_1+1)}/{(3 a_1 + 2 a_2)}}$  in (\ref{solution-riemann-einstein}) and scale $M$
are combined to yield Newton's constant $G_N$. This is the gravitational field equations in GR in that it is based on the metric tensor
$g_{\alpha\beta}$ and its connection $\gamma^{\mu}_{\alpha\beta}$ defined in (\ref{LC}) such that all possible sources -- matter, radiation, cosmological constant -- 
are generically included in $T_{\alpha\beta}$.

Now, we give a brief discusssion of the results found so far. In the beginning there was only the affine geometry \cite{affine} nestling the action (\ref{action}). 
The metric tensor and hence the Levi-Civita connection were born dynamically in (\ref{soln2-riemann}) to satisfy the equation of motion (\ref{soln-riemann}). 
Accompanying these geometrical structures is the non-geometrical tensor field ${\mathcal{T}}^{\mu}_{\alpha\nu\beta}$. Looking at (\ref{soln4-riemann}), 
one may get the impression that ${\mathcal{T}}^{\mu}_{\alpha\nu\beta}$ provides a source for Riemann tensor itself, and hence, generalizes Einstein 
field equations to a full-fledged rank-four tensor equation. This, however, is not true. The reason is that (\ref{soln4-riemann}) is not by itself 
a solution of the equations of motion (\ref{eom-riemann}). Namely, the coefficients $\tilde{c}_i$ are determined not from (\ref{soln2pp-riemann}) alone 
but from (\ref{soln-riemann}), and their solutions in (\ref{coeffs-riemann}) do indeed involve the parameters $c_1$, $a_1$, $a_2$ appearing in (\ref{soln-riemann}). 
Consequently, the Einstein field equations (\ref{eq-einstein}) are all one can get from the Riemann-only action density (\ref{lagran}). Addition of more 
terms to (\ref{lagran}) cannot change this conclusion.

In the present formalism, energy-momentum tensor arises from contractions of a higher-rank tensor field. It is similar to
obtaining Ricci tensor from Riemann tensor but there is a big difference in that ${\mathcal{T}}^{\mu}_{\alpha\nu\beta}$ is a non-geometrical
tensor field, and the identities (\ref{diverg0}) and (\ref{diverg}) it enjoys are different than Bianchi identities. Nonetheless,
deriving energy-momentum tensor from a higher-rank tensor field provides a useful framework. The reason is the cosmological
constant problem. Indeed, energy-momentum tensor of matter and vacuum, when convoluted by graviton propagator in curved background,
defines a connection, the so-called stress-energy connection \cite{demir1}, such that the associated curvature tensor (say, ${\mathcal{T}}^{\mu}_{\alpha\nu\beta}$ above)
involves vacuum energy not as the cosmological constant but as the gravitational constant. In other words, $T_{\alpha\beta}$
in (\ref{eq-einstein}) can be lacking the cosmological term. This mechanism \cite{demir1,demir2,nima}, constructed with no
explicit action principle, canalizes vacuum energy towards gravitational constant, and hence, ameliorates
the cosmological constant problem. 

In the present formalism, matter sector reveals itself through the energy-momentum tensor, only. Given
thus energy density, momentum density, momentum flux, shear stress and pressure of matter, the energy-momentum
tensor can be constructed with no need to an explicit matter action \cite{en-mom}. Despite this, an
action functional is needed for quantization of matter and  determination of conserved charges. However,
in affine spacetime, one cannot write a matter action like (\ref{action}) because there is not any
natural determinant to represent matter fields. The way out from this controversy is that
matter action emerges dynamically along with the metric tensor, connection and energy-momentum tensor.
In other  words, once metric and energy-momentum tensors are known, matter action is constructed as that
functional of matter fields whose variation with respect to the metric tensor gives the energy-momentum tensor.

Let us now have a group-theoretic discussion of the present formalism. As is obvious from (\ref{action}), we have 
started with a spacetime  $(\mathbb{M}, \Gamma)$ governed by an affine connection $\Gamma$ only \cite{affine}. The absence
of a metric disables one from defining an origin as well as having notions of lengths and angles. At this point, the action 
density (\ref{action}) satisfies the symmetry of the diffeomorphism group $\mathrm{Diff} (\mathbb{M})$ on the manifold 
such that the manifold becomes an affine space instead of a vector space \cite{oktay1}. In other words, the affine space $\mathbb{M}$ 
is identified with the quotient group
\begin{eqnarray}
 \mathrm{Diff} (\mathbb{M}) / \mathrm{Diff} (\mathbb{M})_x^1
\end{eqnarray}
where $\mathrm{Diff} (\mathbb{M})_x^1$, a normal subgroup of the diffeomorphism group, is the point 
stabilizer of $x$, where one recalls that a point stabilizer group fixes a point $x$ on an affine 
space providing a vector space and a metric tensor.  If the point stabilizer is the general linear 
group $\mathrm{GL} (1,3)$ then the group of diffeomorphisms $\mathrm{Diff}(\mathbb{M})$ would be the affine group.

However, after solving the equations of motion (\ref{eom-riemann}) and fixing the density factor of the solution (\ref{soln-dav}) and (\ref{M-tensor}), 
we have intrinsically realized a breakdown of the diffeomorphism group into its subgroup $\mathrm{Diff}(\mathbb{M})_x^1$ such that
 \begin{eqnarray}
  \mathrm{Diff} (\mathbb{M}) \leadsto \mathrm{Diff}(\mathbb{M})_x^n \times G_\psi
 \end{eqnarray}
where $n$ denotes a representation of the subgroup on subject, and $G_\psi$ is the group of the internal symmetries which the 
matter fields $\psi$ enjoy {\it e. g.} the Standard Model gauge group. Under this reduction the action decomposes to read
 \begin{eqnarray}
 \label{eq:the_new_action}
  S [\Gamma] \leadsto S [g] + S[g, \psi]
 \end{eqnarray}
such that energy-momentum tensor $\mathbb{T}_{\mu\nu} (\psi)$ is nothing but the Goldstone fields of the broken symmetry.
 
More precisely, by varying the action (\ref{action}) and imposing the appropriate symmetries together with the determinant fixed 
as in (\ref{M-tensor}), the solution (\ref{solution-riemann-einstein}) of the equations of motion provides  two rank-2 tensor fields 
$g_{\alpha\beta}$ and $\mathbb{T}_{\alpha\beta}$ such that the affine connection uniquely splits to yield
\begin{eqnarray}
 \Gamma_{\alpha\beta}^\lambda = \gamma_{\alpha\beta}^\lambda (g) + \Delta_{\alpha\beta}^\lambda (g, \mathbb{T})
\end{eqnarray}
where $\Delta_{\alpha\beta}^\lambda (g, \mathbb{T})$ is a rank-$(1,2)$ tensor field, which consists of the Goldstone degrees of 
freedom for the matter sector, 
and $\gamma_{\alpha\beta}^\lambda$ is a connection compatible with the tensor field $g_{\alpha\beta}$. As a result, 
$\gamma$ and $g$ are nothing but the Levi-Civita connection and the metric tensor, respectively. Consequently, the broken 
$\mathrm{Diff}(\mathbb{M})_x^n$ naturally comprises the Poincar\'{e} group $O(1,3) \rtimes \mathbb{R}^{1,3}$ 
which is indeed a subgroup of the affine group. Note that a breakdown to larger groups like inhomogeneous conform-affine group $\mathrm{CA}(1,3) \rtimes \mathbb{R}^{1,3}$ or the inhomogeneous conform-orthogonal group $\mathrm{CO}(1,3) \rtimes \mathbb{R}^{1,3}$ where conformal transfomations get involved would give rise to a richer phenomenology since the tensorial connection $\Delta_{\alpha \beta}^\lambda$ happens to be a dynamical quantity consisting of non-metricity (See, \cite{demir1, demir2, Karahan:2012yz} for explicit constructions of this tensor field, and \cite{Ali:2007hu,oktay1,oktay2} for further considerations).

The results and discussions above have firmly established that the Riemann-only actions like (\ref{action})
correctly reproduce the Einstein field equations in the presence of matter. Stating explicitly, ignoring the higher-rank origin of
energy-momentum tensor in (\ref{eq-einstein}), the Riemann-only action $I\left[\Gamma\right]$ in (\ref{action}) is equivalent to
\begin{eqnarray}
\label{ac-equiv}
I\left[\Gamma\right]  \leadsto I_{GR}\left[g,\psi\right] = \int d^{4}x \sqrt{-g} \left\{ \frac{1}{16 \pi G_N}  g^{\alpha \beta} {\mathcal{R}}_{\alpha\beta}\left(\gamma\right) +
{\mathcal{L}}_m\left(g,\psi\right) \right\}
\end{eqnarray}
which is nothing but the Einstein-Hilbert action governing the GR. This action completely reproduces the Einstein field equations
(\ref{eq-einstein}) from $\displaystyle{\frac{\delta I_{GR}} {\delta g^{\alpha\beta}} = 0}$ by using
$\displaystyle{T^{\alpha\beta} = 2 \frac{\delta {\mathcal{L}}_m}{\delta g_{\alpha\beta}} + g^{\alpha\beta} {\mathcal{L}}_m}$
where ${\mathcal{L}}_m\left(g,\psi\right)$ is Lagrangian of the matter fields, whose motion equations are given
$\displaystyle{\frac{\delta I_{GR}} {\delta \psi} = 0}$. In general, the equivalence in (\ref{ac-equiv}) is deformed
when $T_{\alpha\beta}$ does not contain the cosmological constant since this situation is known to arise through higher-rank
origin as implemented in \cite{demir1,demir2,nima}. In this case, the equivalent action (\ref{ac-equiv}) obtains non-local
structures deriving from, for instance, an appropriate operator $G_N\left(\Box\right)$ replacing Newton's constant.

Einstein field equations have been formulated also in equivalent Lagrangian approaches. In \cite{AK}, for instance,
it has been shown that  affine space Lagrangians involving the Ricci tensor and matter fields give rise to 
the Einstein field equations (see also \cite{other-AK}). The symmetric and antisymmetric parts of the Ricci 
tensor give way to the metric tensor and a vector field, respectively. The matter is present from the scratch. 
Though this equivalence formally works fine, it is difficult to incorporate known matter actions into affine space 
because there is no analogue of the curvature determinants in the matter sector. As critically discussed in \cite{wilczek},
for scalar fields, for instance, construction of determinants necessitates quartic derivative terms, and 
the resulting action can hardly contain the known interactions.

The Riemann-only affine formalism introduced and analyzed in this work provides an alternative route to Einstein field equations.
It has the potential of naturally accommodating higher-curvature approaches to cosmological constant problem. It can, in general,
be useful in circumstances where the grand curvature of spacetime is under focus. For instance, recent attempts \cite{strings} for introducing
generalized geometries for strings end up with vacuum Einstein equations in 20 dimensions (formed by spacetime and winding degrees of 
freedom of strings). These generalized geometries can be analyzed from the present perspective to follow the grand curvature of spacetime
fully. Concluding briefly and plainly, we state that Einstein field equations for gravitation uniquely derive from gravitational
actions which involve only the Riemann tensor, and this new approach can bring physical insight in different  applications.

\end{document}